\documentclass[a4paper,11pt]{article}
\pdfoutput=1
\usepackage{jcappub}

\usepackage{amsmath}
\usepackage{bm}
\usepackage{xcolor,color,colortbl}
\usepackage[utf8]{inputenc}
\usepackage{hyperref}
\usepackage{url}
\usepackage{graphicx}
\usepackage{multirow,bigdelim}
\usepackage{amssymb}
\usepackage[amssymb]{SIunits}
\usepackage{multirow}
\usepackage{bbold}
\usepackage{verbatim}

\definecolor{darkgreen}{HTML}{339933}

\title{Optical depth to reionization from perturbative 21cm clustering
}

\author[a,b,c]{Noah Sailer,}
\author[a,b]{Shi-Fan Chen}
\author[a,b,c]{and Martin White}
\affiliation[a]{Department of Physics, University of California, Berkeley, CA 94720, USA}
\affiliation[b]{Berkeley Center for Cosmological Physics, UC Berkeley, CA 94720, USA}
\affiliation[c]{Lawrence Berkeley National Laboratory, One Cyclotron Road, Berkeley, CA 94720, USA}

\emailAdd{nsailer@berkeley.edu}
\emailAdd{shifan\_chen@berkeley.edu}
\emailAdd{mwhite@berkeley.edu}

\abstract{The optical depth $\tau$ is the least well determined parameter in the standard model of cosmology, and one whose precise value is important for both understanding reionization and for inferring fundamental physics from cosmological measurements.
We forecast how well future epoch of reionization experiments could constraint $\tau$ using a symmetries-based bias expansion that highlights the special role played by anisotropies in the power spectrum on large scales.  Given a parametric model for the ionization evolution inspired by the physical behavior of more detailed reionization simulations, we find that future 21cm experiments could place tight constraints on the timing and duration of reionization and hence constraints on $\tau$ that are competitive with proposed, space-based CMB missions provided they can measure $k\approx 0.1\,h\,\text{Mpc}^{-1}$ with a clean foreground wedge across redshifts spanning the most active periods of reionization, corresponding to ionization fractions $0.2 \lesssim x \lesssim 0.8$.
Significantly improving upon existing CMB-based measurements with next-generation 21cm surveys would require substantially longer observations ($\sim5$ years) than standard $\mathcal{O}(1000 \,\,\text{hour})$ integration times.
Precise measurements of smaller scales will not improve constraints on $\tau$ until a better understanding of the astrophysics of reionization is achieved.  In the presence of noise and foregrounds even future 21cm experiments will struggle to constrain $\tau$ if the ionization evolution deviates significantly from simple parametric forms.
}

\begin{document}
\maketitle
\flushbottom

\section{Introduction}
\label{sec:introduction}

Within the Standard ($\Lambda$CDM) Model of Cosmology the optical depth to reionization $\tau$ is the least well-determined of the standard parameters \cite{PCP18}.  Though it is an almost purely astrophysical parameter its uncertainty has a large impact on inferences about fundamental physics because, when fitting cosmic microwave background (CMB) data, it is strongly degenerate with parameters like the amplitude of primordial fluctuations ($A_s$, $\sigma_8$) and the sum of the neutrino masses $M_\nu$. In addition to hindering higher-precision measurements of these parameters, $\tau$ derives from the Thomson scattering of free electrons and is thus a probe of the history of cosmic reionization \cite{Becker15,Mesinger16,McQuinn16}. Beyond suppressing the amplitude of CMB temperature anisotropies this Thomson scattering results in a range of physical phenomena in the CMB including large-scale (E-mode) polarization \cite{PlanckLegacy18} and the kinetic Sunyaev-Zeldovich (kSZ) effect \cite{SZ80,Dvorkin:2008tf,Ferraro:2018izc,Alvarez:2020gvl} that can be used to measure $\tau$. In addition, probes of the Epoch of Reionization (EoR) have the potential to directly constrain the global ionization history and therefore the abundance of free electrons that source $\tau$ \cite{Barkana05,McQuinn06,Liu:2015txa, billings2021extracting,Greig:2020suk,Greig:2020hty}. Our goal in this paper is to investigate this final method using recent advances combining cosmological perturbation theory with techniques based on effective field theories.

There have been numerous approaches to modeling the 21cm signal from the EoR, ranging from detailed radiative transfer simulations coupled to hydrodynamics \cite{Pawlik17,Kannan22} to semi-analytic \cite{Munoz22} and hybrid \cite{Trac21} models.  The majority are based on N-body simulations.  By contrast, our approach to modeling 21cm observations will be based upon cosmological perturbation theory with a symmetries-based bias expansion.  This approach is now quite commonplace in large-scale structure studies, but differs in some key respects from the standard approaches within the reionization field (see references above for examples of analytic or semi-numerical models and refs.~\cite{Hoffman19,McQuinn:2018zwa,Qin22} for notable examples similar in spirit to our approach).  No matter how complex the physical processes leading to ionized regions, as long as they obey certain symmetries (e.g.\ rotational invariance and the equivalence principle) and are quasi-local we can write $\delta_{21}$ as an expansion in powers of the density field, tidal field and low-order derivatives \cite{McDRoy09,Des16}.  This expansion will be valid on large scales, in our case on scales larger than the typical size of an ionized bubble \cite{McQuinn:2018zwa,Qin22}.  The impact of small-scale physics not explicitly included in the model is then handled by correction terms (i.e.\ counterterms), which are also restricted by symmetry considerations.  By marginalizing over the bias parameters and counterterms we can focus on the cosmological information contained in large scale modes while efficiently removing sensitivity to small-scale physics.  Since our expansion is based entirely on symmetry, it makes minimal assumptions about that physics.  Conversely, information about the sources and morphology of reionization is encapsulated in the small-scale modes and bias coefficients.

Given that perturbation theory and the bias expansion marginalize away much of the ``astrophysics'' of the 21cm signal it is worth asking whether it is still possible to recover useful information about reionization within such a framework, or whether the study of reionization necessarily requires detailed modeling of the small-scale physics of the EoR (e.g.\ the composition of ionizing sources). For example, while the overall amplitude of the observed 21cm power spectrum is proportional to the brightness temperature $\overline{T}_b$ and can therefore be used to predict the electron fraction (\S\ref{ssec:21cm_clustering}), even at linear order in perturbation theory the power spectrum will also be proportional to the square of the linear bias $b^2$ whose value cannot be determined absent detailed knowledge of small-scale behavior of neutral hydrogen, the host halos of ionizing sources etc., leading to an exact degeneracy with the electron fraction.

\begin{figure}
\centering
\includegraphics[width=0.99\linewidth]{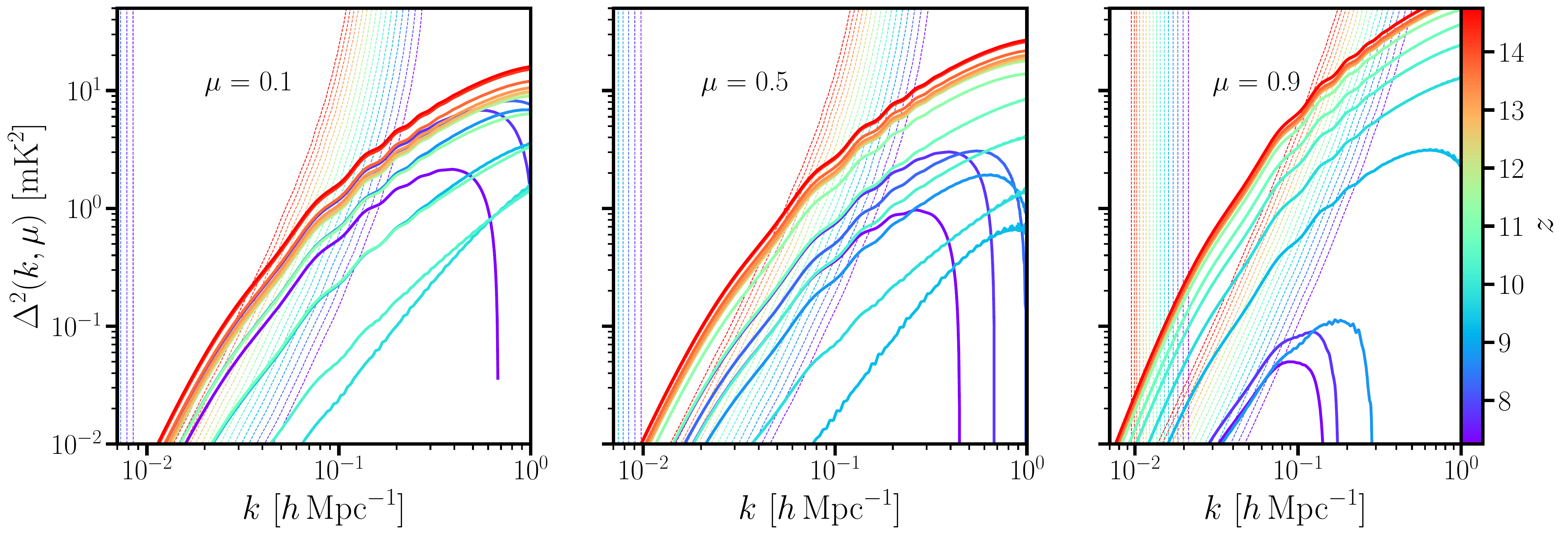}
\caption{The dimensionless 21cm power spectrum ($\Delta^2 = k^3P/2\pi^2$) as a function of redshift for our fiducial model (solid lines, see \S\ref{sec:fiducial_scenario}) and the thermal noise per mode (steeply rising, dashed lines) for the case of 1000 14m dishes with a 10 year integration time (first column of Table \ref{tab:forecasts}). The step in thermal noise (vertical lines) at low-$k$ is due to a absence of baselines capable of measuring modes with $k_\perp \chi/2\pi < D/\lambda_{21}^\text{obs}$, where $D$ is the diameter of the dishes. The panels show $\mu=0.1$ (left), $\mu=0.5$ (middle) and $\mu=0.9$ (right) with colors indicating redshift.  Note that the spectrum is quite anisotropic (the amplitudes vary significantly with $\mu$, particularly at lower redshifts), with the shape encoding valuable information about the cosmology and reionization history. 
}
\label{fig:power_spectra}
\end{figure}

However, while many potential measurements of the global reionization history, e.g.\ the overall amplitude of the 21cm signal, are polluted by small-scale physics, the presence of fundamental symmetries implies that others are protected. For example, the equivalence principle dictates that all matter behaves identically under the force of gravity, such that the peculiar velocity of neutral hydrogen on large scales must be the same\footnote{The relative velocity between baryons and dark matter is discussed in \S\ref{sec:conclusions}.} as that of matter and other tracers like galaxies. Since 21cm surveys infer line-of-sight distances in terms of observed frequencies these peculiar velocities are then encoded in anisotropies in the observed clustering due to redshift-space distortions (RSD), where an object at $\bm{x}$ appears to be at redshift-space position $\bm{s} = \bm{x} + \bm{u}$, with $\bm{u} = \hat{\bm{n}}\ (\hat{\bm{n}} \cdot \bm{v}_{\rm pec}) / \mathcal{H}$. Within linear theory the redshift-space power spectrum is given by \cite{Kai87}
\begin{equation}
    P_{\textrm{obs}}(k) = \overline{T}_b^2 (b + f \mu^2)^2 P_{\rm lin}(k), \quad \mu = \hat{\bm{n}} \cdot \hat{\bm{k}}
\end{equation}
where $\hat{\bm{n}}$ is the line of sight unit vector and $f(z) = d\ln D/d\ln a$ is the growth rate, with $f \simeq 1$ to sub-percent precision at the matter dominated redshifts where reionization occurs (see Fig.~\ref{fig:power_spectra}). As the growth rate and $P_{\rm lin}$ are purely a function of cosmology, the size of the anisotropy ($\mu^2$) on large scales is then uncontaminated by small-scale physics and for any given cosmology can be used to infer the brightness temperature $\overline{T}_b$, unlike the isotropic piece proportional to $b^2$ \cite{Barkana05,McQuinn06}. In addition to the linear bias, $\overline{T}_b$ is perfectly degenerate in linear theory with the amplitude of matter fluctuations $\sigma_8$ within a single redshift bin. This degeneracy can be broken\footnote{At lower redshifts non-linear evolution of large-scale structure can be used to break this degeneracy \cite{Castorina19,Sailer:2021yzm}, but at the early times relevant for EoR such effects are very small on the scales that can be reliably modeled.} using the different time-evolution of $\sigma_8(z) \propto (1+z)^{-1}$ and $\overline{T}_b(z) \propto (1+z)^{1/2}[1-x(z)]$, or by combing 21cm observations with primary CMB fluctuations (which constrain $\sigma_8 e^{-\tau}$).

\begin{figure}
\centering
\includegraphics[width=0.48\linewidth]{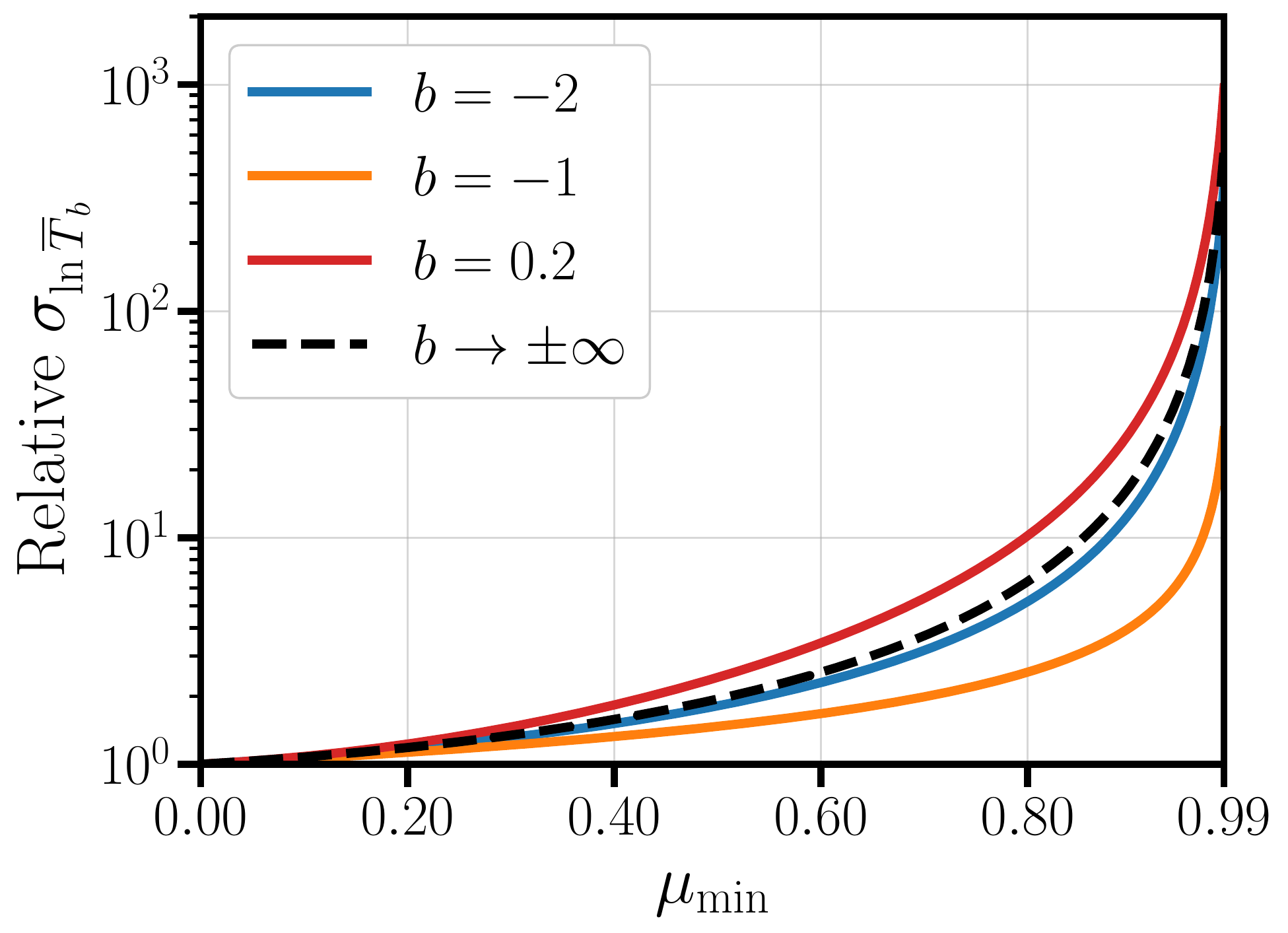}
\includegraphics[width=0.49\linewidth]{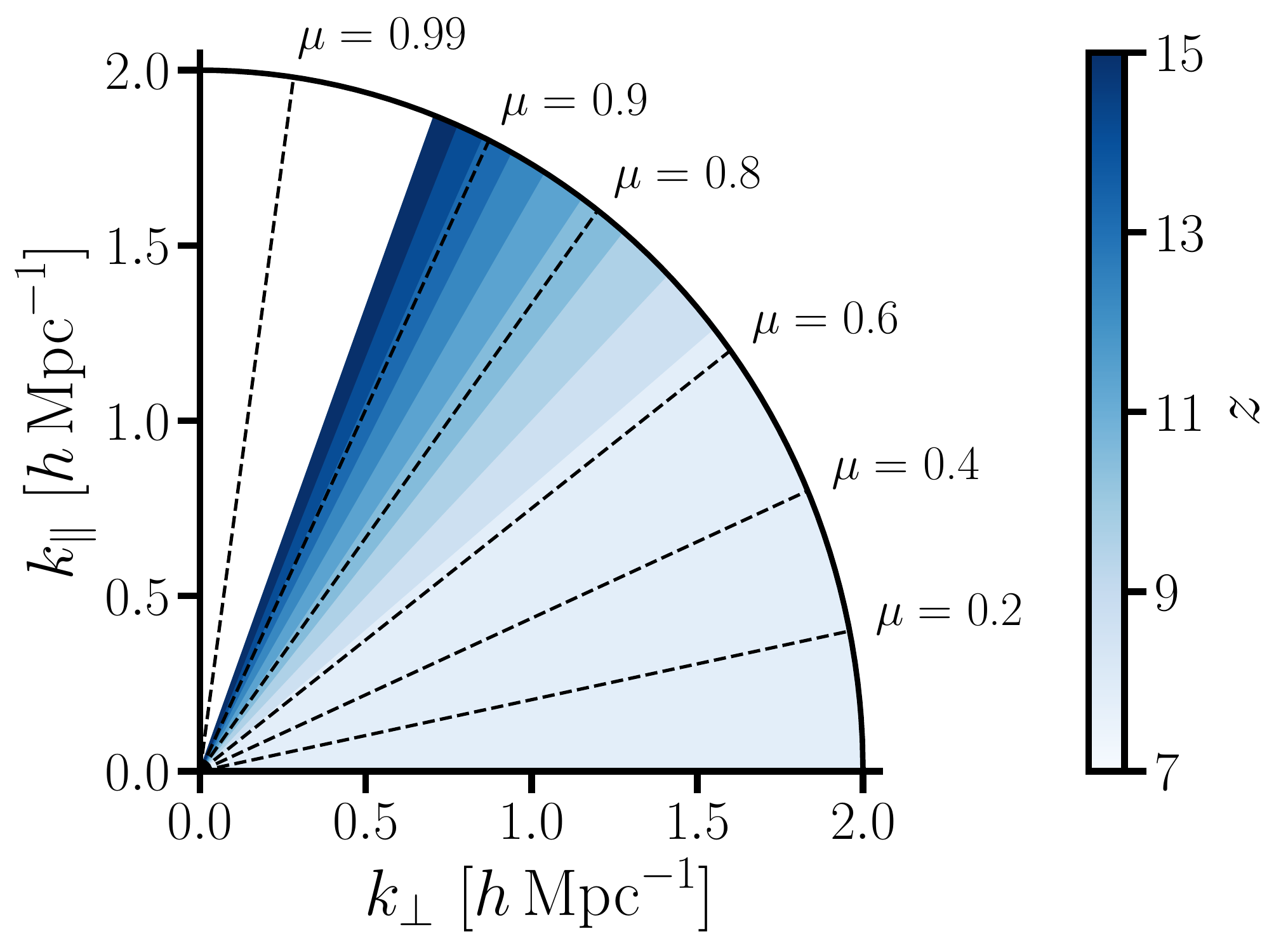}
\caption{
\textit{Left:} Error on $\ln \overline{T}_b$ (normalized to $\mu_{\rm min}=0$) after marginalizing over the linear bias, as a function of $\mu_\text{min}$, following the simplified scenario outlined in \S\ref{sec:introduction}. The colors denote the assumed fiducial value of the linear bias.
\textit{Right:} Foreground wedge for our fiducial scenario ($N_w=3$; Eq.~\ref{eqn:wedge}) as a function of redshift, assuming $14\,$m diameter dishes (see \S\ref{sec:realism} for a more detailed discussion of the foreground wedge).  }
\label{fig:intuition}
\end{figure}

The left panel of Figure~\ref{fig:intuition} shows our relative ability to measure the brightness temperature as a function of the minimum angular scale $\mu_{\rm min}$ in an idealized setup\footnote{Specifically, we consider a scenario where the survey is noiseless and the signal is given by linear theory, and the treat the linear power spectrum $P_\text{lin}$ as fixed.  Our ability to constraint $\overline{T}_b$ is then independent of the linear power spectrum shape and the Fisher matrix is given by
\begin{equation*}
    F_{ij} \propto \int_{\mu_\text{min}}^1 d\mu\ {\left[\partial_i (b+\mu^2)^2 \overline{T}^2_b\right] \left[\partial_j (b+\mu^2)^2 \overline{T}^2_b\right]}
    \bigg/
    {(b+\mu^2)^4 \overline{T}^4_b},
\end{equation*}
where $i \in \{\overline{T}_b, b\}$.
}. Clearly, since our ability to constrain $\overline{T}_b$ depends on resolving the anisotropy due to RSD, reducing the angular range over which we fit results in dramatic reductions in constraining power (see also ref.~\cite{Pober15}). This is particularly relevant for 21cm interferometers wherein chromaticity in instrument leads to foreground leakage into a ``wedge'' below some $\mu_{\rm min}$ that renders those modes unusable (\S\ref{sec:realism}). The wedge angle corresponding to our fiducial setup is shown in the right panel of Figure~\ref{fig:intuition}. 

Our goal in this paper will be to investigate the feasibility of using the EoR 21cm signal in redshift space to constrain the global reionization history within a perturbation theory framework. The remainder of this paper is organized as follows:
Section \ref{sec:tau} discusses the relationship between the optical depth and reionization as probed by 21cm measurements.  This section describes our parameterizations of the ionization fraction evolution.
Section \ref{sec:LPT} gives a quick overview of Lagrangian perturbation theory and the symmetries-based bias expansion that we employ throughout this work.
Section \ref{sec:forecasts} discusses the Fisher matrix formalism that we employ in our forecasts, and the fiducial models and scale cuts that we investigate.
We conclude in Section \ref{sec:conclusions} with a discussion of implications from our forecasts.

\section{Extracting the optical depth from 21cm clustering}
\label{sec:tau}

In this section we describe our method for constraining $\tau$ from a measurement of 21cm clustering during the reionization era.  We start by identifying the physical quantities that we need to constrain in order to infer $\tau$ and then describe how these are related to the 21cm power spectrum before discussing how we choose to parameterize the evolving ionization fraction.

\subsection{The optical depth as a derived parameter}

We begin by reviewing how the optical depth can be computed given a cosmological background and a reionization history (i.e.\ the evolution of the ionization fraction) \cite{Liu:2015txa}. The mean CMB optical depth is by definition $\tau = \sigma_T\int_0^{z_\text{rec}} dz\,\bar{n}_e(z) dr/dz$, where $\sigma_T$ is the Thompson cross-section, $\bar{n}_e(z)$ is the volume-averaged free electron (proper) number density, $z_\text{rec}\simeq 1100$ is redshift of last scattering, and $dr/dz$ is the line-of-sight (proper) distance per unit redshift. Throughout we assume that the Universe is flat, in which case $dr/dz = c/(1+z)H(z)$, and use overlines to denote \textit{volume} averages. 
We assume that free electrons are sourced solely from hydrogen and (singly ionized) helium, neglecting the free electrons from doubly-ionized helium\footnote{Ref.~\cite{Liu:2015txa} performs a back-of-the-envelope calculation to show that doubly-ionized helium contributes to $\sim 1.4\%$ of the total optical depth. Assuming that the best fit $\tau=0.0561$ from Planck is the truth, this amounts to $\tau_\text{HeIII}\sim 0.0008$.} and additional isotopes of H and He, as well as heavier elements. We further assume that hydrogen and helium have identical ionization fractions: $x_\text{HII} \equiv x_\text{HeII}$. This is a standard approximation in the literature \cite{Madau97,Clesse:2012th,Liu:2015txa}, due to the similar ionization potentials (13.6 vs $24.6\,$eV) of the two elements and the fact that neutral H efficiently shields the He from ionizing radiation.
Given these assumptions, the free electron number density can be rewritten as $n_e(\bm{x}) = x_\text{HII}(\bm{x}) \rho_b(\bm{x})/\mu m_\text{H}$, where $\rho_b$ is the baryon density, $\mu=1+(m_\text{He}/m_\text{H}-1)Y_p/4 $ is the mean molecular weight and $Y_p = 4 n_\text{He}/n_b$ is the primordial helium abundance. It is standard to rewrite the baryon density as $\rho_b(\bm{x}) \equiv \overline{\rho}_b(z)[1+\delta_b(\bm{x})]$, where $\overline{\rho}_b(z) = 3H_0^2 \Omega_b(1+z)^3/8\pi G$ is the mean baryon density, and $\delta_b$ is the density contrast. With these definitions and assumptions in hand, the optical depth can be rewritten as \cite{Liu:2015txa}
\begin{equation}
\begin{aligned}
\label{eq:tau}
    \tau &=  \frac{3 H_0^2 \Omega_b \sigma_T c}{8\pi G \mu m_\text{H}} \int_0^{z_\text{rec}} dz\,\frac{(1+z)^2}{H(z)}
    \overline{x_\text{HII}(1+\delta_b)}(z)
    \\
    &= 0.0691 \frac{\omega_b}{h\mu}
    \int_0^{z_\text{rec}} dz\,\frac{(1+z)^2}{E(z)} x(z),
\end{aligned}
\end{equation}
where $\omega_b = \Omega_b h^2$, $H_0 \equiv 100\, h\, \text{km}/\text{s}/\text{Mpc}$, $E(z) = H(z)/H_0$, and we've defined the \textit{mass-weighted} average ionization fraction $x(z) \equiv \overline{x_\text{HII}(1+\delta_b)}(z)$. In going from the first to second line of Eq.~\eqref{eq:tau} we have substituted the values of quantities which are known to well beyond the precision of our $\tau$ measurement. In particular we see that $\tau$ may be derived from the baryon fraction $\omega_b$, the expansion history $E(z)$, the primordial helium abundance $\mu = 1 + 0.7428\,Y_p$, and the ``reionization history'' $x(z)$.

\subsection{Ionization fraction from 21cm clustering}
\label{ssec:21cm_clustering}

The 21cm brightness temperature, defined as the difference between the observed 21cm brightness and that of the CMB, at redshift-space position $\bm{s}$ is given by \cite{2012MNRAS.422..926M,Furlanetto:2006jb}:
\begin{equation}
\label{eq:Tb}
\begin{aligned}
    T_b(\bm{s},z)
    &= 
    \frac{3}{16}
    \frac{\hbar c^3 A_{10}}{k_B \nu_{21}^2}
    \frac{[1-x_\text{HII}(\bm{x},z)]n_\text{H}(\bm{x},z)}{(1+z)(dv_\parallel/dr_\parallel)}\left[1- \frac{T_\text{CMB}(z)}{T_s(\bm{x},z)}\right]
    \\
    &\equiv 
    \overline{T}_b(z)[1+\delta_{21}(\bm{s},z)]
\end{aligned}
\end{equation}
where $\bm{x}$ is the corresponding position in real space, $\langle \delta_{21}\rangle=0$, $A_{10} = 2.85 \times 10^{-15}\,\,\text{Hz}$
is the spontaneous emission coefficient of the 21cm transition, $\nu_{21} = 1.42\,\,\text{GHz}$ is the rest frame 21cm transition frequency and $T_s$ is the spin temperature, defined so that the ratio of the number of excited (spin aligned) hydrogen atoms to the number of hydrogen atoms in the ground state (spins anti-aligned) is given by $3\exp[-2\pi\hbar\nu_{21}/k_B T_s]$\footnote{This expression follows from the more general form
\begin{equation}
    T_b(\textbf{s},z) = \frac{T_s - T_{\rm CMB}}{1+z} (1 - e^{-\tau_{\rm 21}(\textbf{s})})
\end{equation}
where $\tau_{\rm 21}$ is the 21cm optical depth, which in the small $\tau_{\rm 21}$ limit gives $1 - e^{-\tau_{\rm 21}} \approx \tau_{\rm 21}$. Since the 21cm signal comes from the diffuse, optically-thin IGM this approximation is expected to be a good one, but it is worth noting that in the opposite, optically thick limit the nonlinearity of the exponential mapping tends to contaminate the linear redshift-space signal critical to obtaining $\tau$ constraints. This is due to the fact that the exponential mapping breaks the number conservation symmetry that protects the Kaiser form of RSD \cite{Chen21}.
}. The derivative $dv_\parallel/dr_\parallel$ is the (proper) gradient of the (proper) velocity along the line of sight and sources the redshift space anisotropy in the observed signal; dividing a real-space density by this factor yields the redshift-space density.
We assume that the spin temperature dominates over the CMB $(T_s \gg T_\text{CMB})$, and discuss the limitations of this approximation in \S\ref{sec:realism}. Under these assumptions, the mean 21cm brightness temperature takes the form:
\begin{equation}
    \overline{T}_b(z)
    =
    189.1[1-x(z)]\frac{(1+z)^2}{E(z)}
    \ \frac{\omega_b}{h\mu}
    \left(1-\frac{Y_p}{4}\right) \quad\mathrm{mK} ,
\end{equation}
where we have substituted the values of known quantities. 

A 21cm interferometer directly measures the Fourier modes $T_b(\bm{k})$, from which one measures the power spectrum:
\begin{equation}
    P_\text{obs}(\bm{k},z) = \overline{T}_b^2(z) P_{21}(\bm{k},z) + P_N(\bm{k},z),
\end{equation}
where $\langle \delta_{21}(\bm{k})\delta_{21}(\bm{k}')\rangle=(2\pi)^3\delta^D(\bm{k}+\bm{k}')P_{21}(\bm{k})$ and $P_N(\bm{k},z)$ is the thermal noise of the instrument. In particular, we note that the mean 21cm brightness temperature controls the overall amplitude of the observed (redshift space) power spectrum of the 21cm signal. This mean temperature is in turn proportional to the neutral fraction $1-x$. Thus, by measuring the amplitude of the 21cm power spectrum one can in turn infer a measurement of the ionization fraction $x$.

\subsection{Parametric reionization history}
\label{sec:parametric_reio_hist}

The presence of a series of acoustic peaks in the angular power spectra of the CMB indicates that the Universe was dense and ionized at early times and then  underwent a rapid transition to being (largely) neutral at $z\,{\simeq}\,1100$.  This neutral period lasted for a significant time before the first astrophysical sources (in currently popular models, massive stars in relatively low mass early galaxies) reionized the Universe.  Lower limits on the redshift of reionization come from the lack of complete absorption of UV photons from high redshift QSOs \cite{Gunn65}.  While significant questions remain, there are numerous observational constraints on the process of reionization \cite{Becker15,Mesinger16,McQuinn16}.  Current observations point towards a ``late and fast'' reionization period, though with considerable uncertainty \cite{Zhu2022}.  

Rather than use a simulation-based model to infer the evolution of the ionization fraction from a set of physical parameters (star formation rates, escape fractions, baryon content, etc.), we will instead take a more astrophysics-agnostic approach and adopt an analytic parameterization of $x(z)$. Ref.~\cite{Trac21} found that the Weibull function, which has three free parameters:
\begin{equation}
    x(z)
    =
    \text{Exp}
    \left[
    -\text{max}
    \left(
    \frac{z-a}{b},0
    \right)^c\,
    \right],
\end{equation}
is capable of accurately fitting a wide range of reionization scenarios. We will adopt this Weibull form as the default model in our forecasts. A comparison of the best-fit Weibull function with our fiducial reionization history is shown in the left panel of Fig.~\ref{fig:fiducial_stuff}. 

In \S\ref{sec:forecasts} we will consider how freeing up the functional form of $x(z)$ impacts our constraints. To examine this we will model $x(z)$ with Lagrange interpolating functions, as proposed by ref.~\cite{Trac:2018jla}. The parameters of the model are a set of mass-weighted ionization fractions $\{x(z_i)\}$ for a chosen (fixed) set of redshifts $\{z_i\}$. These points are smoothly connected via:
\begin{equation}
    x(z) 
    =
    \text{min}\left[
    \exp\left(
    \sum_{i=1}^N p_i(z)
    \right),1\right]
    \quad
    \text{where}
    \quad
    p_i(z) \equiv \ln(x(z_i)) \prod_{j\neq i}
    \frac{\ln(1+z)-\ln(1+z_j)}{\ln(1+z_i)-\ln(1+z_j)}.
\end{equation}
We consider how our constraints vary as a function of the number of free parameters in \S\ref{sec:z_independent_cuts}.

\section{The 21cm clustering signal in Lagrangian Perturbation Theory}
\label{sec:LPT}

To robustly account for scale-dependent bias and structure formation in the quasi-linear regime we will adopt the formalism of Lagrangian Perturbation Theory (LPT). The literature on perturbative modeling of large-scale structure is vast and we will not try to give a comprehensive overview of the subject or rederive results; rather, our goal is to give a reasonably self-contained summary and highlight important physical points distinguishing the 21cm signal from the more standard treatment of galaxies.  A review of perturbation theory can be found in refs.~\cite{Ber02,Dodelson20}, and of bias in ref.~\cite{Des16}.

Within LPT cosmological structure formation is modeled as the gravitational dynamics of an effective fluid whose short-wavelength (UV) physics are integrated out, with corresponding effects on large-scale structure parametrized by effective-theory counterterms and stocastic contributions \cite{VlaWhiAvi15,PorSenZal14}. Unlike in Eulerian effective theories of large-scale structure (see e.g.\ \cite{BNSZ12,CHS12}), which model the resulting fluid via its densities and velocity potentials, within the Lagrangian picture of fluid dynamics one traces the trajectories of infinitesimal fluid elements labeled by their initial (Lagrangian) positions $\bm{q}$. The dynamics are encoded in a displacement vector $\bm{\Psi}(\bm{q},\eta)$ defined such that the Eulerian (comoving) positions $\bm{x}$ of the fluid elements at a conformal time $\eta$ is $\bm{x}(\bm{q},\eta) = \bm{q} + \bm{\Psi}(\bm{q},\eta)$, 
and where the displacement vector is sourced by the gravitational potential to follow $\ddot{\bm{\Psi}}(\bm{q})+\mathcal{H}\dot{\bm{\Psi}}(\bm{q}) = -\nabla_{\bm{x}} \Phi(\bm{x})$, with $\mathcal{H}$ defined as the conformal Hubble parameter. The gravitational potential in turn follows Poisson's equation, with the matter (over)density obtained via number conservation:
\begin{equation}
\label{eq:overdensity}
    1 + \delta_\mathrm{m}(\bm{x},\eta) 
   = \int d^3\bm{q}\ \delta^D(\bm{x}-\bm{q}-\bm{\Psi}(\bm{q},\eta)).
\end{equation}
The velocity field is given straightforwardly as the time derivative of the displacement.

In general, cosmological observables like the 21cm signal do not directly probe the matter field described by the displacements $\bm{\Psi}$ but rather are determined by complex small-scale astrophysical phenomena and their response to the large-scale matter distribution over cosmic time \cite{Des16}. For example, whether a galaxy exists at a spacetime point $(\bm{x},\eta)$ will generically depend on the matter density, velocity gradients and tidal fields in a neighborhood about its trajectory $\bm{q} + \bm{\Psi}(\bm{q},\eta)$ approximately the size of its Lagrangian halo radius $R_h$. While the size of the nonlocality in the galaxy bias expansion is governed primarily by $R_h$ (since the dominant physical effect is the nonlinear gravitational collapse of the Lagrangian patch into a halo), the 21cm signal during reionization is sensitive to other, stronger nonlocalities such as the mean free path $\lambda_{\rm mfp}$ of ionizing photons, which unlike nonrelativistic matter can travel significant ($\sim H^{-1}$) distances in a Hubble time in the absence of absorption. In this case the size of nonlocalities is roughly set by the size of the ionized bubbles during reionization, with smaller bubbles earlier on in the process corresponding to our ability to perturbatively model the signal over a wider range of scales \cite{McQuinn:2018zwa}.

Within the (Lagrangian) effective-theory framework \cite{McDRoy09,Sen14,Ass14,VlaCasWhi16,Chen20a} we can write the quasi-local dependence of the 21cm signal at $\bm{x}$ on neighboring points along its trajectory --- that is, nearby in $\bm{q}$ --- as an expansion order-by-order in the initial conditions at $\bm{q}$
\begin{align}
\label{eq:bias_expansion}
  F(\bm{q}) = 1 + b^\text{L}_1 \delta(\bm{q}) + \frac{1}{2}b^\text{L}_2 \big(\delta^2(\bm{q})-\langle\delta^2\rangle\big) &+ b^\text{L}_s \big(s^2(\bm{q})-\langle s^2\rangle \big) + \cdots \nonumber \\
  & + R_\ast^2 \nabla^2 \delta(\bm{q}) + \cdots + \epsilon(\bm{q}).
\end{align}
This functional is then dynamically advected to the real-space position $\bm{x}$ \cite{Mat08b}:
\begin{equation}
    1 + \delta_{\rm 21}(\bm{x}) 
   = \int d^3\bm{q}\,F(\bm{q})\, \delta^D(\bm{x}-\bm{q}-\bm{\Psi}(\bm{q}))
\label{eqn:delta21}
\end{equation}
in the same way as the matter field (Eq.~\ref{eq:overdensity}).\footnote{Fluctuations in the 21cm signal can be sourced by fluctuations in either the baryon density $\delta_b$ or the electron fraction $\delta_{x}$ and treating these separately is common in 21cm studies. However, since it can equally be thought of as fluctuations in the density of neutral hydrogen, in this work we directly model this product using a single bias expansion instead of separately.  This avoids needing to model correlated fluctuations in $\delta_b$ and $\delta_{x}$ when evaluating $\delta_b\delta_{x}$, which in perturbation theory language is known as a ``composite operator''.} In this way the 21cm clustering signal is captured in LPT via a combination of dynamics ($\bm{\Psi}$) and dependence on the initial conditions $F(\bm{q})$.

The form of the functional $F(\bm{q})$ is restricted by symmetries and the equivalence principle to depend only on scalar combinations of the density, velocity gradients and the tidal field. Within perturbation theory these quantities can all be written in terms of the initial density field, i.e.\ to linear order scalars formed from $\partial_i v_j$ and $\partial_i \partial_j \Phi \sim (k_i k_j / k^2) \delta$.  To second order, contributions to  the bias expansion can be summarized by its dependence on $\delta^2$ and $s^2 = s_{ij} s_{ij}$, where $s_{ij}\equiv ( \partial_i \partial_i/\partial^2 -  \delta_{ij}/3 )\ \delta$ is the shear field. 
Beyond the local value of these initial conditions along the trajectory starting at $\bm{q}$, the existence of nonlocalities due to e.g.\ ionized bubbles can be handled by including derivative corrections such as $\nabla^2\delta$, i.e.\ performing a Taylor series expansion of $\delta$ around $\bm{q}$. These corrections become increasingly prominent as we probe scales closer to the bubble size.  Depending on how small in scale we need to fit to it may be necessary to include terms higher order in density and derivatives, though it is important to stress that our perturbative treatment of $F(\bm{q})$ as a quasi-local functional will necessarily break down on scales smaller than the bubble size where the physics of reionization are nonlocal.

The last term in Eq.~\ref{eq:bias_expansion} is the residual, $\epsilon(\bm{q})$, which represents stochastic small-scale contributions to the clustering of the 21cm signal uncorrelated with the large-scale modes we will be interested in. To lowest order these modes only correlate in close proximity $\langle \epsilon(\bm{q}) \epsilon(\bm{q}')\rangle \sim \delta^D(\bm{q}-\bm{q}')$ or, equivalently, have a flat spectrum in Fourier space: $P_{\epsilon}(k) = \textrm{const}$.

Finally, 
we need to account for contributions to the \textit{observed} densities due to peculiar velocities in the form of redshift-space distortions (RSD). Specifically, since line-of-sight distances are inferred from redshifts they incur an additional Doppler contribution from the peculiar velocity $\bm{u} = \hat{\bm{n}}\ (\hat{\bm{n}} \cdot \bm{v}_{\rm pec}) / \mathcal{H}$, such that the 21cm signal in redshift-space is given by \cite{Mat08a,VlaWhi19,Chen20c}
\begin{equation}
    1 + \delta_{\rm 21}(\bm{s}) 
   = \int d^3\bm{q}\,F(\bm{q})\, \delta^D(\bm{s}-\bm{q}-\bm{\Psi}_s(\bm{q})) \quad ; \quad \bm{\Psi}_s(\bm{q}) = \bm{\Psi} + \bm{u}.
\label{eqn:delta21_rsd}
\end{equation}
In the limit where the real-to-redshift-space mapping is unique ($\bm{s} \leftrightarrow \bm{x} \leftrightarrow \bm{q}$) we can integrate over the delta function to recover the familiar expression $\delta(\bm{s}) = aH/ (dv_\parallel/dr) \delta(\bm{x})$, where $v_\parallel$ is the line of sight velocity including Hubble flow \cite{Furlanetto:2006jb}.
Within the Einstein-de Sitter approximation, which is an excellent approximation at the redshifts where reionization occurs, the $m^{\rm th}$ order displacement obeys the simple relation
\begin{equation}
    \bm{\Psi}^{(m)}_s(\bm{q},\eta) = \bm{\Psi}^{(m)} + m f \hat{\bm{n}}\ \left(\hat{\bm{n}} \cdot \bm{\Psi}^{(m)}\right) .
\end{equation}
with $f\simeq 1$, though it is important to note that, like the real-space densities, the velocities too require effective-theory corrections to properly account for their dependence on small-scale physics \cite{Chen20a}. To lowest order, the above calculation recovers the familiar Kaiser \cite{Kai87} form:
\begin{equation*}
    1 + \delta_{21} = \frac{1 + b^L_1 \delta(\bm{q})}{1 + (aH)^{-1}\, dv^\text{pec}_\parallel/dr}
\end{equation*}
and, using that $v_{\textrm{pec}, i}/(aH) = - f \nabla^{-1}_i \delta$ we have in Fourier space
\begin{equation*}
    \delta_{21}(\bm{k}) = ([1 + b^L_1] + f \mu^2)\, \delta(\bm{k}), \quad \mu = \hat{\bm{n}} \cdot \hat{k}
\end{equation*}
such that the linear theory power spectrum is given by $(b + f\mu^2)^2 P_{\rm lin}(k)$ \cite{Kai87}, where $b = 1 + b_1^L$.  This is the familiar result used in the introduction.

\section{Forecasts}
\label{sec:forecasts}

Now we turn to our forecasts of how well a (future) 21cm interferometer could determine $\tau$ by measuring the evolving 21cm power spectrum.  Our forecasts use the Fisher matrix formalism, with an extensive set of nuisance terms to handle the complexities of the reionization process.  These forecasts allow us to determine the key observations that would allow a robust extraction of $\tau$ and highlight critical assumptions.  We begin by defining our fiducial cosmology, reionization history and 21cm clustering model before describing the range of scales and redshifts used in our fits.

\subsection{Model and fiducial scenario}
\label{sec:fiducial_scenario}
In this work we use $\verb|CLASS|$ \cite{CLASS} to compute the linear CDM+baryon power spectrum, and $\verb|velocileptors|$\footnote{\href{https://github.com/sfschen/velocileptors}{https://github.com/sfschen/velocileptors}} \cite{Chen20a} to model non-linearities and redshift-space distortions (see \S\ref{sec:LPT}) to one-loop order within LPT:
\begin{equation}
\label{eqn:stochastic}
    P_{21}(\bm{k})
    =
    P_{21}^\text{1-loop}(\bm{k})+
    \big(\alpha_0 +\alpha_2 \mu^2 + \alpha_4 \mu^4\big) k^2 P^\text{Zel}_{cb}(\bm{k})
    + N_0 + N_2(\mu k)^2 + N_4(\mu k)^4,
\end{equation}
where $N_2$ and $N_4$ encode the effects of small-scale velocities, $N_0$ is the shot noise, $P^\text{Zel}_{cb}$ is the CDM+baryon power spectrum in the Zel'dovich approximation, and $\mu=\hat{\bm{k}}\cdot\hat{\bm{n}}$ is the cosine of the angle between the wavevector and the line of sight. This model includes a bias expansion to quadratic order in the linear CDM+baryon density contrast (i.e.\ the first line of Eq.~\ref{eq:bias_expansion}). For further details on the calculations see refs.~\cite{Chen20a,Chen20c}.

We use the $\verb|FishLSS|$\footnote{\href{https://github.com/NoahSailer/FishLSS}{https://github.com/NoahSailer/FishLSS}} \cite{Sailer:2021yzm} code to compute the Fisher information for the full-shape 21cm power spectrum $P_\text{obs}(\bm{k})$ in the basis:
\begin{equation}
\label{eq:basis}
(A_s, n_s, h, \omega_c, \omega_b,M_\nu) \cup \{\ln\overline{T}_b,b, b_2, b_s, \alpha_0, \alpha_2, \alpha_4, N_0, N_2, N_4\}_z,
\end{equation}
where $M_\nu$ is the sum of the neutrino masses, which we hold fixed unless otherwise stated, and the biases here $(b,b_2,b_s)$ are in the Eulerian frame. We treat all time-dependent nuisance terms (the right-hand set in Eq.~\ref{eq:basis}; the $z$ subscript indicates time-dependence) in different redshift bins as separate, independent variables which are individually varied. We direct the reader to ref.~\cite{Sailer:2021yzm} for additional details regarding the numerical computation of derivatives and evaluation of the full shape Fisher matrix, as well as information regarding the CMB Fisher matrices used below. Unless otherwise stated we will include a prior on $\Lambda$CDM+$M_\nu$ from Planck primary CMB measurements in all of our forecasts.

As discussed in \S\ref{sec:parametric_reio_hist} we will parameterize the reionization history with set of parameters $\{x_i\}$. Using this parametrization we can trade the free parameters $\{\ln\overline{T}_b\}_z$
for the set $\{x_i\}$ via the chain rule: $\partial_{x_i}P_\text{obs}(z_j) = (\partial_{x_i} 
\ln\overline{T}_b)\partial_{\ln\overline{T}_b} P_\text{obs}(z_j)$, since the value of the ionization fraction has little impact on the evolution of large-scale matter fluctuations and primarily impacts $P_\text{obs}$ through an amplitude change. In our fiducial scenario we use the Weilbull form ($\{x_i\} = \{a,b,c\}$), in which case our fiducial full set of free parameters is given by
\begin{equation}
\label{eq:fiducial_basis}
    \underbrace{(A_s, n_s, h, \omega_c, \omega_b,a,b,c)}_{\equiv\,\bm{\theta}}
    \cup \{b, b_2, b_s, \alpha_0, \alpha_2, \alpha_4, N_0, N_2, N_4\}_z.
\end{equation}
In this scenario our total number of free parameters is $8 + 9\times(\text{number of redshift bins})$. 
Once we have the Fisher matrix in the basis Eq.~\eqref{eq:fiducial_basis}, we then average over the nuisance parameters $\{b, b_2, b_s, \alpha_0, \alpha_2, \alpha_4, N_0, N_2, N_4\}_z$ to obtain the nuisance-averaged $8\times 8$ Fisher matrix $\bm{F}$, and use Eq.~\eqref{eq:tau} to propagate the errors on $\bm{\theta}$ to the optical depth: $\sigma_\tau^2 = \partial_{\bm{\theta}}\tau\cdot\bm{F}^{-1}\cdot\partial_{\bm{\theta}}\tau$. 

\begin{figure}[!h]
\centering
\includegraphics[width=0.46\linewidth]{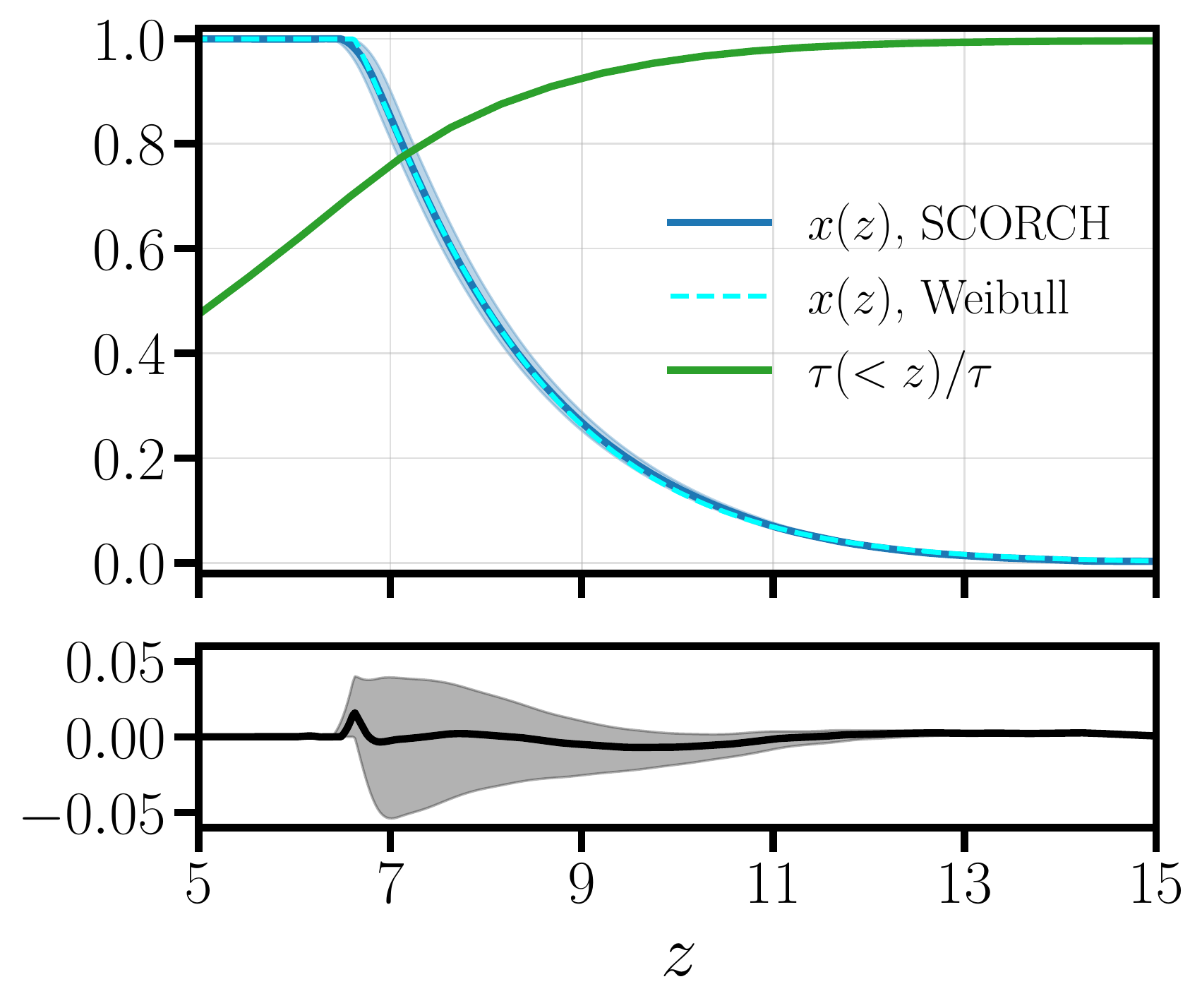}
\includegraphics[width=0.46\linewidth]{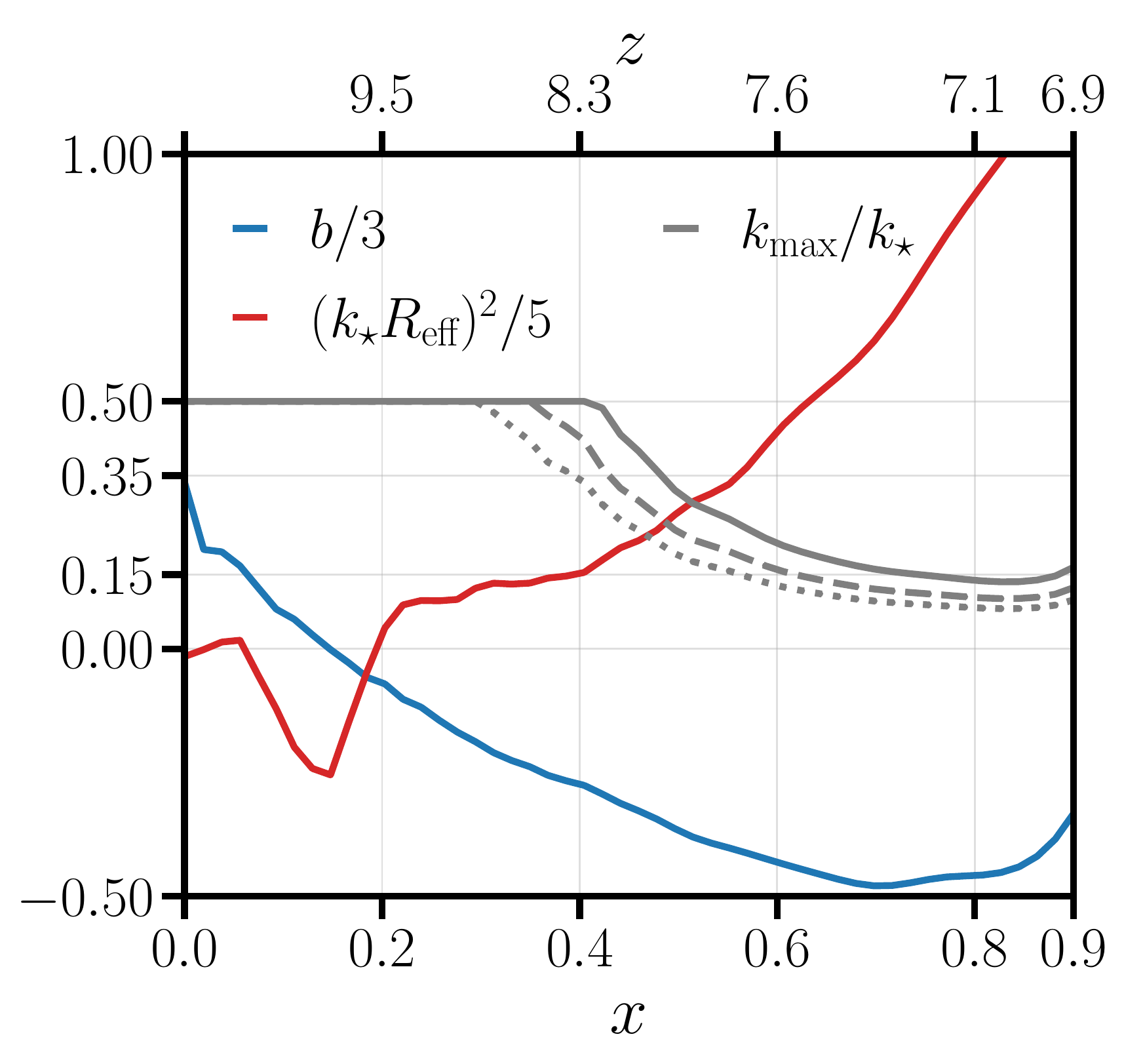}
\caption{
\textit{Left}: In the top panel we show our fiducial reionization history (blue) and the best fit Weibull function (cyan). The bottom panel shows the residual $|\delta x| < 0.02$. This residual results in a systematic error of $|\delta \tau| = 0.0005$. The edges of shaded bands represent the $\delta z$ by which one can shift the reionization history and change the optical depth by $|\delta \tau| = 0.001$. In green we show the (relative) cumulative contribution to the optical depth for redshifts $<z$. 
\textit{Right}: The fiducial linear Eulerian bias $b$, the effective bubble size $R_\text{eff}$ and a representative evolution of $k_\text{max}$ (see Eq.~\eqref{eq:kmaxz}), where $k_\star = 1\,h\,\text{Mpc}^{-1}$. The grey solid, dashed and dotted lines correspond to $X_\alpha = 3,\,4,\,5$ respectively.  Such limits ensure that for $k<k_{\rm max}$ the corrections are perturbative and the higher-order counterterms are negligible. 
}
\label{fig:fiducial_stuff}
\end{figure}

We base our fiducial reionization history off of the SCORCH \cite{Doussot2017} simulations, which generalize the  radiation-hydrodynamic code $\verb|Radhydro|$ \cite{Trac2004,Trac2006} to include varying radiation escape fractions: $f_\text{esc}(z) = f_8[(1+z)/9]^{a_8}$. We choose to adopt the redshift-independent scenario ($a_8=0$) as our fiducial model for $x(z)$, which is shown in the left panel of Fig.~\ref{fig:fiducial_stuff}. This scenario predicts $\tau = 0.0639$ when all of the cosmological parameters are fixed at their best-fit Planck values \cite{PlanckLegacy18}, which is within $1.1\sigma$ of the Planck measurement $\tau = 0.0561 \pm 0.0071$.
In Fig.~\ref{fig:fiducial_stuff} we also show the best fit Weibull function ($a=6.62,\,b=1.85,\,c=1.14$), which agrees with the simulation output to within $|\delta x| \leq 0.02$, resulting in a very small systematic shift in the optical depth: $|\delta \tau| = 0.0005$.  Additionally Fig.~\ref{fig:fiducial_stuff} shows the cumulative contributions to $\tau$ from redshifts $<z$ (normalized to the total $\tau$).  Note that $\approx 0.5$ of the total $\tau$ comes from redshifts below $5$ where the Universe is fully ionized (and hence has no uncertainty from ignorance of $x$).  Most of the rest comes from $z<11$.

For the clustering we adopt the evolution of $b$ measured from simulations\footnote{These values are taken from the left-hand panel of Fig.~7. We fit a Weibull function to the reionization history of ref.~\cite{McQuinn:2018zwa}, finding $(a,b,c)=(6.73, 1.0, 0.92)$, and use this to appropriately translate these curves to our fiducial reionization history (assuming that $b(x)$ and $R_\text{eff}^2(x)$ are fixed).} in ref.~\cite{McQuinn:2018zwa} and set our fiducial counterterm $\alpha_0(z)$ from the effective bubble radius evolution $R_{\rm eff}(z)$ in the same work (Fig.~\ref{fig:fiducial_stuff}). These biases are consistent with those measured by ref.~\cite{Qin22}, however other theoretical calculations have found different values \cite{Xu:2019pcz}. The effective bubble radius\footnote{The measured values of $R^2_\text{eff}$ are poorly constrained at early times \cite{McQuinn:2018zwa}, so the negative values of $R^2_\text{eff}$ may just be a numerical artifact.  This has a negligible impact on our constraints since in this regime the hard $k^h_\text{max}$ cut dominates over the bubble size such that $\alpha_0 k^2\ll 1$.} is defined such that $\delta_{21} \supset b (1 - R_\text{eff}^2 k^2/3)\delta$, so that the power spectrum $P_{21}\supset  -2b^2 R_\text{eff}^2 k^2 P/3 + \cdots$, and so we identify $\alpha_0 =  - 2 b^2 R_\text{eff}^2 /3$. For the higher order biases, $b_2$ and $b_s$, we assume vanishing Lagrangian bias corresponding to the coevolution values for Eulerian biases and we further take $N_0=N_2=N_4=\alpha_2=\alpha_4=0$.  Since these terms are relevant primarily at small scales and in the absence of receiver noise, the precise values are not important. 
We have explicitly checked that changing the fiducial value of $b_2$ from its coevolution value to $b_2(z)=-2$, a value that roughly agrees with that measured by ref.~\cite{McQuinn:2018zwa} from simulations, impacts our forecasts (for the idealized scenario of \S\ref{sec:z_independent_cuts}) by less than 7\%.

In Fig.~\ref{fig:fiducial_stuff} we note that the linear bias starts off positive and eventually becomes negative while $R_{\rm eff}$ grows towards the end of reionization.  The growth of the bubbles is self-explanatory.  Ref.~\cite{McQuinn:2018zwa} gives a simple explanation for the evolution of the bias assuming that the ionized bubbles trace the locations of the sources in proportion to their ionizing luminosity and that recombinations and light travel delays can be neglected.  In this model a large region containing more sources will have more ionization so $\delta_{\rm HII}=\delta_S+\delta$, where $\delta_S$ is the source overdensity field.  Writing $\delta = (1-x)\delta_{\rm HI} + x\delta_{\rm HII}$ we note that the first term on the rhs is $\delta_{21}$ while the last is $x(\delta+\delta_S)$ and thus $\delta_{21}=(1-x)\delta-x\delta_S=(1-x[1+b_S])\delta$.  Assuming the sources of reionization are highly clustered we expect $b$ to be negative once reionization gets underway.

\subsection{Idealistic survey}
\label{sec:z_independent_cuts}

We first examine an idealized noiseless survey which covers half the sky over the redshift range $7 < z < 15$, which we will split into 16 redshift bins\footnote{We have checked that doubling the number of redshifs bins (so that $\Delta z = 0.25$) does not significantly impact our forecasts, suggesting that $\Delta z=0.5$ is sufficient for convergence.} with $\Delta z =0.5$. To investigate the sensitivity of these measurements in the $\bm{k}-z$ parameter space, we will first consider redshift-independent scale cuts, with our fiducial cuts being $k_\parallel^\text{min} = 0.05\,\,h\,\text{Mpc}^{-1}$, $k_\text{max}=0.5\,\,h\,\text{Mpc}^{-1}$ and $\mu_\text{min}=0.7$. A more realistic treatment which accounts for the redshift dependence of the foreground wedge and $k_\text{max}$ will be considered in \S\ref{sec:realism}. For this idealized scenario, and using the best-fit Weibull function to parameterize the reionization history, we find a fiducial error on the optical depth $\sigma_\tau = 0.00016$ with a Planck prior on $\Lambda$CDM, and $0.00026$ without (21cm clustering only).  This illustrates that if reionization proceeds smoothly in redshift and can be parameterized by a few parameters then there is more than enough signal in the clustering to infer $\tau$ with very high accuracy.

\begin{figure}[!h]
\centering
\includegraphics[width=0.29\linewidth]{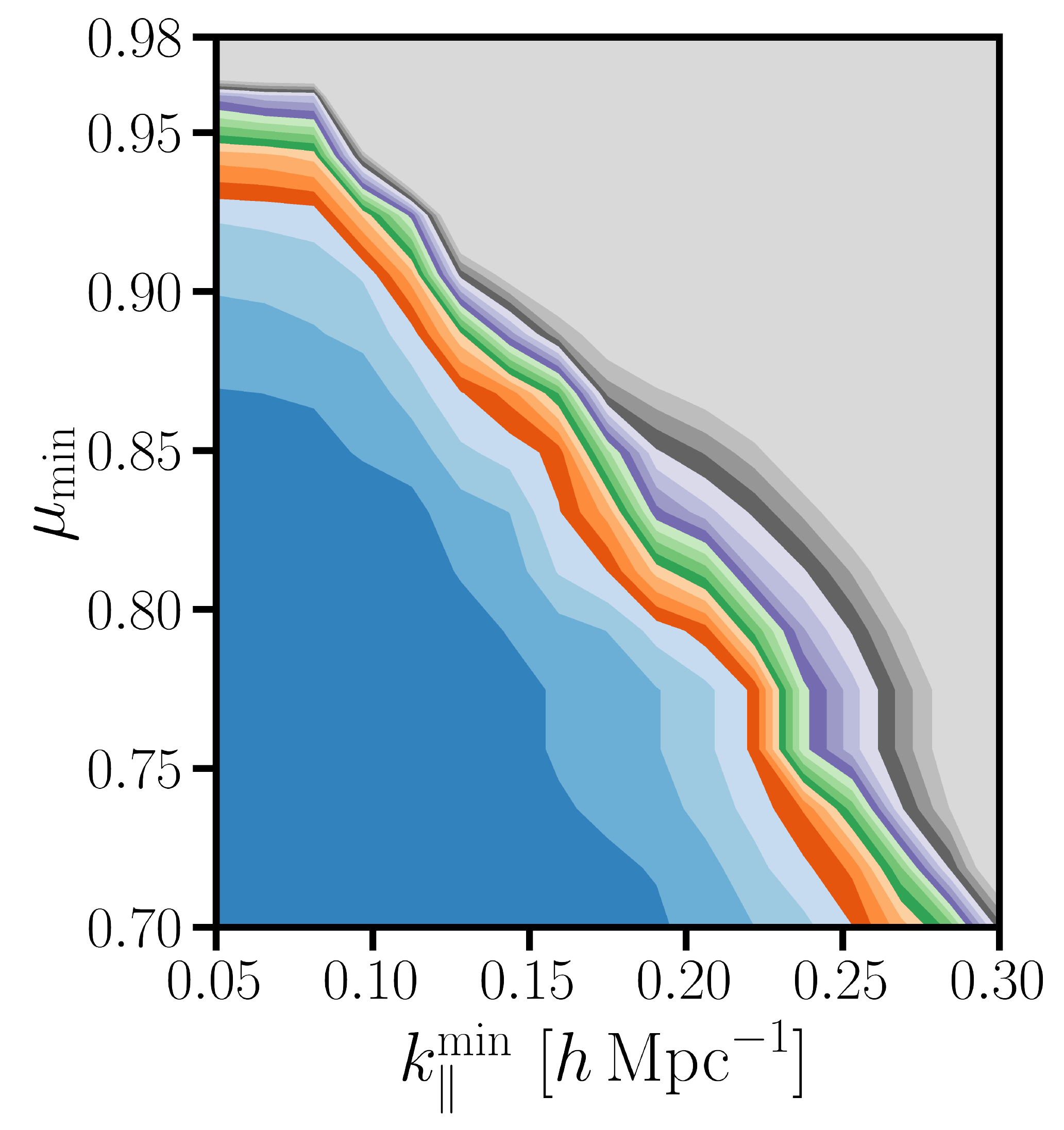}
\includegraphics[width=0.32\linewidth]{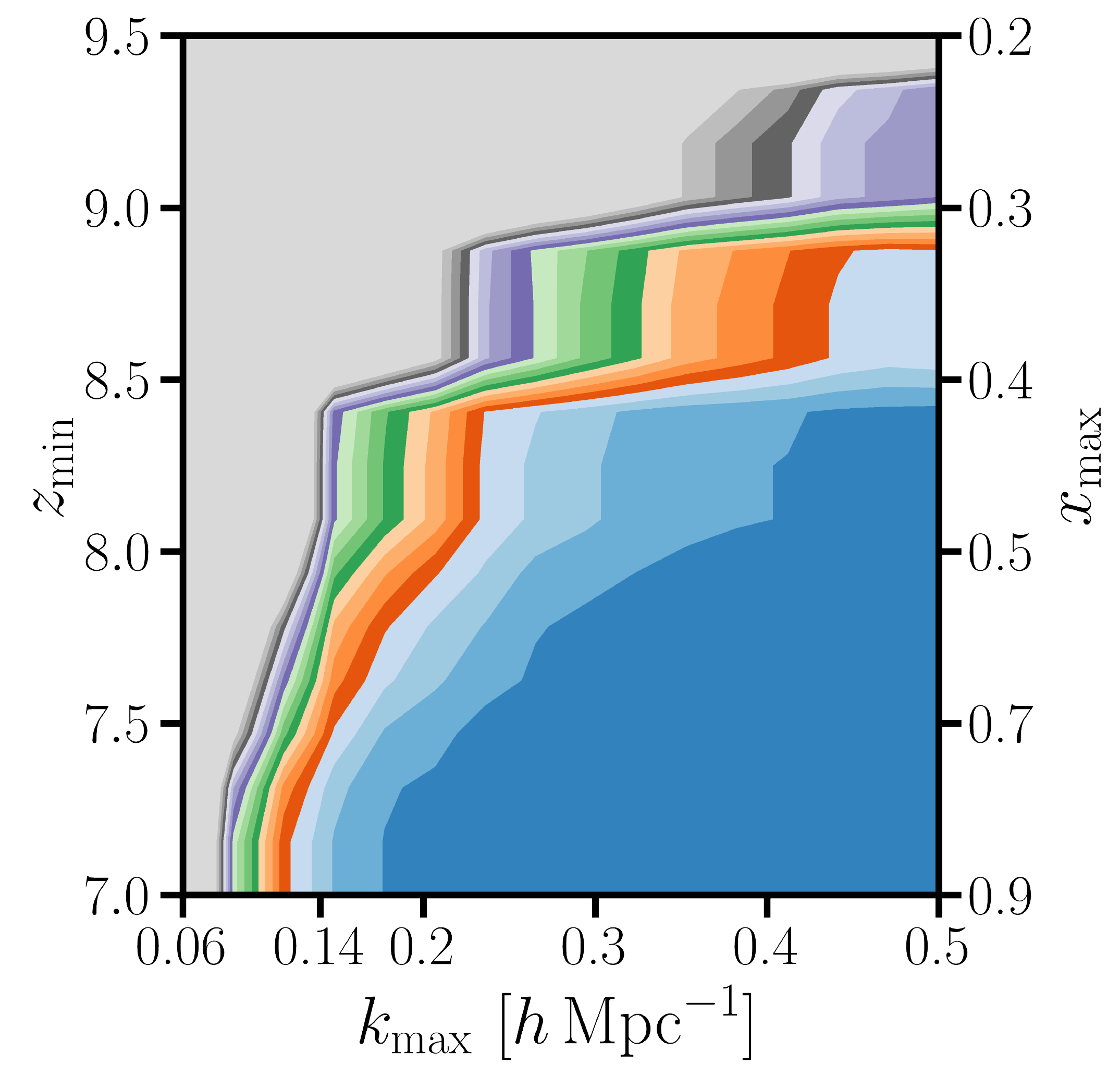}
\includegraphics[width=0.35\linewidth]{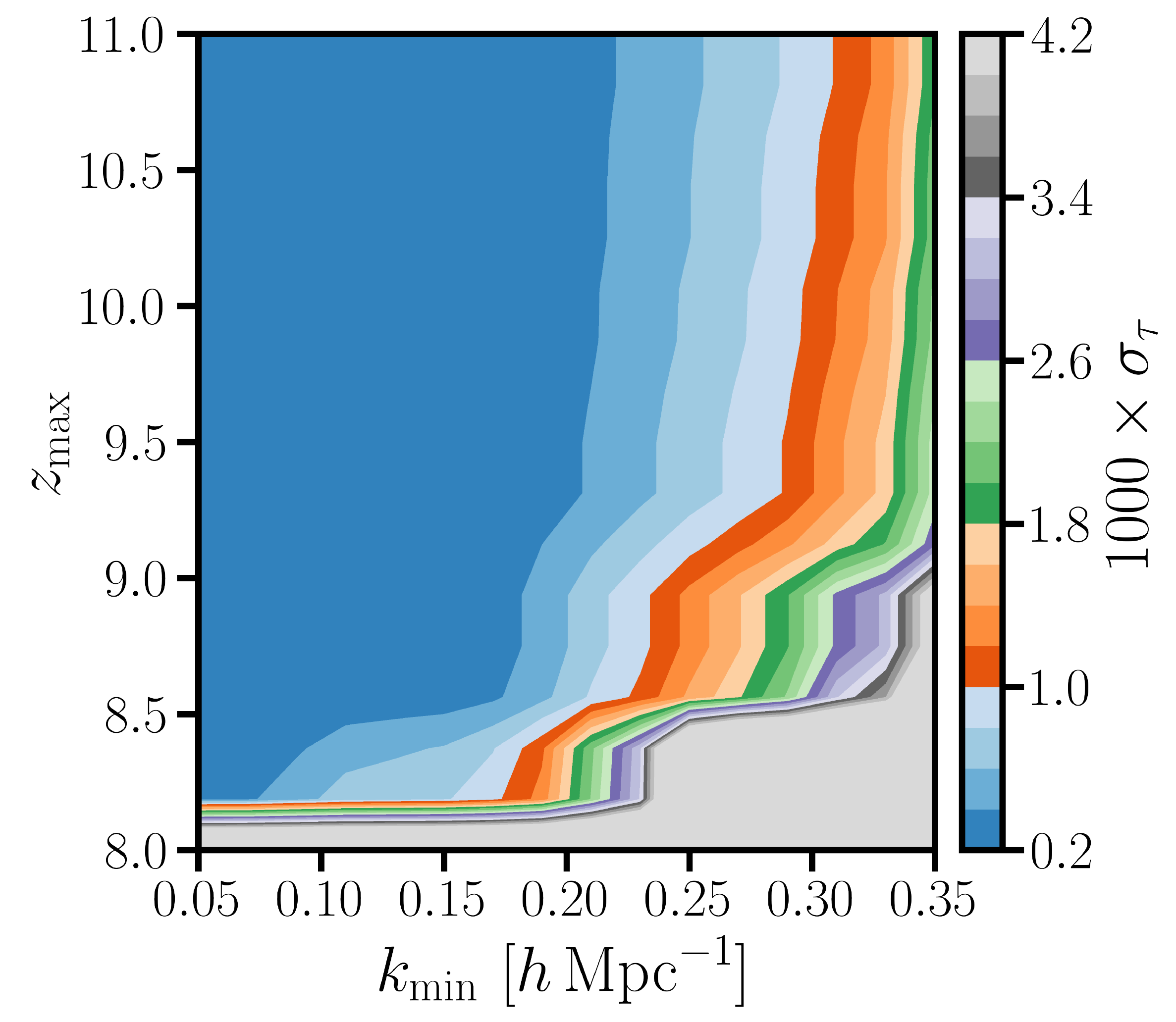}
\caption{Errors on the optical depth (times a thousand) for a noiseless survey covering $7 < z < 15$ over half the sky. The survey has been split into 16 redshift bins with $\Delta z=0.5$. The fiducial scale cuts are $k_\text{max}=0.5\,\,h\,\text{Mpc}^{-1}$, $k_\parallel^\text{min} = 0.05\,\,h\,\text{Mpc}^{-1}$ and $\mu_\text{min}=0.7$, and we include a prior on $\Lambda$CDM from Planck primary CMB measurements.
}
\label{fig:varying_stuff}
\end{figure}

Exploring this further, Fig.~\ref{fig:varying_stuff} shows how the errors depend upon the range of scales and redshifts probed.  As Fig.~\ref{fig:varying_stuff} makes clear, the constraints are more sensitive to an experiment's ability to resolve large scales (i.e.\ low $k$ modes) and obtain sufficient dynamic range in $\mu$ than to the range of redshifts it probes or its ability to probe very small scales.  Among other things this suggests that foreground subtraction and instrument calibration will matter as much for measuring $\tau$ as raw instrumental sensitivity.  Quantitatively we see that the constraints on $\tau$ degrade significantly if the experiment cannot measure $k<0.2\,h\,{\rm Mpc}^{-1}$, but it is only necessary to measure well modes up to $k\simeq 0.2\,h\,\mathrm{Mpc}^{-1}$ to get almost all of the constraining power in the data.  This is because the smaller-scale modes mostly serve to constrain astrophysical parameters.  All of these modes are linear, from the perspective of dynamics.

Much of the integral responsible for $\tau$ comes from `intermediate' redshifts, so the survey does not need to probe very high or low redshifts in order to constrain $\tau$ very well. This is also illustrated in Fig.~\ref{fig:varying_stuff} where we see that the errors on $\tau$ remain small for $z_{\rm min} < 8$ and $z_{\rm max}>9.5$ (in our fiducial reionization history) corresponding to coverage of ionization fractions $0.2\lesssim x(z)\lesssim 0.5$.  Only once the survey is unable to probe those epochs does the inferred error on $\tau$ rise dramatically.  Varying the Weibull parameter $b$, which affects the duration of reionization, in our fiducial model within upper limits on the duration of reionization from the CMB ($b<2.8$) changes our forecasts for $\sigma_\tau$ by $<25\%$.  We expect that $T_S\gg T_\text{cmb}$ for $z\lesssim 10$ \cite{Becker15,Mesinger16,McQuinn16}, justifying our neglect of spin temperature fluctuations in our fiducial model.  If reionization is particularly extended or has a non-trivial high $z$ component this assumption would need to be revisited.  Such scenarios are disfavored, but if the spin temperature can fluctuate while $x$ is significantly different from zero then the connection between 21cm clustering and $\tau$ would be much more complex.

Finally, to investigate how sensitive our constraints are to the number of free variables in our parameterization of the reionization history, we consider swapping the Weibull function for a Lagrangian interpolating function (see \S\ref{sec:parametric_reio_hist}) with $N$ free parameters, which we choose to be the ionization fractions $\{x(z_i)\}$ at redshifts $z_i = 7,7+6/N,\cdots,13$. For $N = 3,\,5,\,7,\,9$ we find $1000\times\sigma_\tau=0.20,\, 0.25,\, 1.95,\,4.47$ for our set of fiducial scale cuts. The scaling of the error with $N$ gets significantly worse with more realistic scale and temporal cuts: for example, after enforcing $z_\text{max} = 10$ and adding a redshift-dependent foreground wedge ($N_w=3$, see \S\ref{sec:realism}) the error increases from 0.4 to 6.1 for $N=3,\,5$.
It is worth noting that not all of the loss of constraining power with increasing $N$ is necessarily physical: astrophysical processes contributing to reionization typically evolve on long (fraction of Hubble) time scales, leading to constraints on the variation of $x(z)$ between interpolating points. Introducing physically motivated smoothness priors to parametrically fit e.g.\ the expansion history have recently been studied in ref.~\cite{Raveri21}, and we leave it to future work to develop these techniques in the context of reionization.

\subsection{More realistic scenarios}
\label{sec:realism}

Having determined what range of scales and redshifts are most important for constraining $\tau$ we now consider more realistic scenarios in which we add instrument noise, foregrounds and redshift evolution.  Our purpose is not to model any particular 21cm survey in detail, but rather to explore the impact that various observational choices have on the inferred constraints on $\tau$.  The current generation of 21cm interferometers does not have sufficient sensitivity to improve upon existing measurements of $\tau$, so we shall consider hypothetical next-generation instruments with integration times on the order of years, consistent with proposed post-reionization 21cm surveys \cite{collaboration2018inflation}, as opposed to the more standard 1000 hours\footnote{Reducing the integration time for our fiducial SKA-like survey (see the SKA-like column in Table \ref{tab:forecasts}) from 5 years to 1000 hours increases the $1\,\sigma$ error on $\tau$ from 0.0009 to 0.0067, which is a marginal improvement over Planck alone ($\sigma_\tau = 0.0071$). For the 1000 dish scenario, reducing the integration time to 1000 hours results in no improvement over Planck in all of our scenarios.}.

Our fiducial instrument will be a hexagonally packed array of $10^3$, non-tracking, $14\,$m dishes. This is a scaled up version of the Hydrogen Epoch of Reionization Array (HERA; \cite{HERA}).  We compute the baseline distribution using $\verb|21cmSense|$\footnote{\href{https://github.com/steven-murray/21cmSense}{https://github.com/steven-murray/21cmSense}} \cite{Pober13b,Pober13c}.  The thermal noise power is based on Appendix D of ref.~\cite{Slosar19a}.  At high redshifts we are sky noise dominated and since most of our constraining power on $\tau$ comes from intermediate $k$ the baseline distribution is of secondary importance and we are mostly sensitive to the total collecting area (and the integration time). HERA frequencies range from 50 to $250\,$MHz, which translates to $4.8<z<27.5$ and we shall assume a future array would cover a comparable frequency range.  As we have seen, such a wide redshift range is more than sufficient to provide tight constraints on $\tau$.

The range of wavenumbers transverse to the line of sight that can be probed by an interferometer depends upon the distribution of baseline separations \cite{TMS17}.  Since the most constraining modes for our purposes are the low $k$ modes we are primarily interested in the number of short baselines.  If we take a minimum dish separation of $14\,$m for our fiducial cosmology $k^\text{min}_{\perp} = 6\times 10^{-3}h\,\mathrm{Mpc}^{-1}$ at $z=10$ and $4\times 10^{-3}h\,\mathrm{Mpc}^{-1}$ at $z=15$.  The baselines that probe $k_\perp \sim 0.1\,h\,\mathrm{Mpc}^{-1}$ are $240\,$m at $z=10$ and $380\,$m at $z=15$.  All of these baselines would be well probed by next-generation experiments. 

The minimum wavenumber that can be probed in the line of sight direction depends upon the ability of the experiment to subtract the foregrounds that are many orders of magnitude brighter than the signal, but spectrally smooth.  Methods for foreground subtraction are rapidly improving, so we shall simply explore how our constraint on $\tau$ depends upon $k_\parallel^\text{min}$, which inherently limits the accessible $\mu$-range. For our fiducial scenario we will take $k_\parallel^\text{min}=0.1\,h\,{\rm Mpc}^{-1}$.  The maximum observationally accessible wavenumber is normally set by thermal noise and the lack of many widely separated baselines.  However since we are attempting to focus on perturbative scales this will not be the only limitation of interest to us.  Rather our $k_\text{max}$ will (also) be set by the scale at which our bias expansion becomes inapplicable, which is mostly set by the evolution of the ``effective'' bubble size.  We shall take $k_\text{max}$ to be the smaller of $k^h_\text{max}=0.5\,h\,\mathrm{Mpc}^{-1}$ and the scale at which our lowest-order counterterm is $X_\alpha^{-2}$ of the linear power spectrum:
\begin{equation}
\label{eq:kmaxz}
    k_\text{max}(z) = 
    \text{min}\left[
    \frac{1}{X_\alpha\sqrt{|\alpha_0(z)|}}\,,\,
    k^h_\text{max}
    \right]
    \quad
    \text{where}
    \quad
    \alpha_0(z) = -\frac{2}{3}b^2(z)R^2_\text{eff}(z)
\end{equation}
and consider $X_\alpha=3$ and 4. In principle the validity of the bias expansion in real space is set strictly by the nonlocality scale $R_{\rm eff}$ but we have adopted $1/\sqrt{\alpha_0}$ to serve as a catch-all, including both the validity of the bias expansion and the significance of the linear RSD signal from which most of our constraint derives; we have checked that this choice does not significantly impact our $\sigma_\tau$ forecast.

Finally, contamination from systematic errors in the inherently chromatic response of an interferometer, interacting with astrophysical sources that are many orders of magnitude brighter than the signals of interest, leads to contamination within a ``foreground wedge'' \cite{Abdurashidova22,Gan22}.  A particularly lucid and pedagogical introduction to the physical origin of the wedge can be found in ref.~\cite{Morales12}.
Given the rapidly evolving state of 21cm analysis, we shall take a simple model of the wedge and explore how variations in performance will affect our inference.  In particular we follow ref.~\cite{collaboration2018inflation} in modeling the foreground wedge as allowing only modes with
\begin{equation}
    k_\parallel 
    >
    \text{max}\left[
    k_\perp
    \frac{\chi(z)H(z)}{c(1+z)} \sin(\theta_w(z))
    \,\,,\,\,k_\parallel^\text{min}\right]
    \quad
    \text{where}
    \quad 
    \theta_w(z) =
    N_w
    \frac{1.22}{2 \sqrt{0.7} } 
    \frac{\lambda_{\rm obs}}{D_\text{phys}}
\label{eqn:wedge}
\end{equation}
where $\lambda_{\rm obs}=(1+z)\,21\,$cm is the observed wavelength, $D_\text{phys}$ is the physical diameter of the dish, $k_\parallel = k\mu$ and $k_\perp = k\sqrt{1-\mu^2}$. For our fiducial scenario we will take $N_w=3$.

Fig.~\ref{fig:intuition} shows the foreground wedge as a function of redshift for $N_w=3$ and $D_{\rm phys}=14\,$m.  The minimum $\mu$ uncontaminated by foregrounds rises from $\simeq 0.5$ at $z= 6$ to $\simeq 0.9$ at $z= 15$.  For comparison the minimum $\mu$ kept in the current analysis of HERA data is $\simeq 0.98$ \cite{Abdurashidova22,Qin22}, which would render measurement of RSD ineffective.  By contrast SKA predicts an observable window of $\mu > 0.67$.  Clearly considerable effort remains to be done on instrument calibration and foreground removal to enable our forecasts.

\newcolumntype{g}{>{\columncolor{gray!20}}c}
\begin{table}[!h]
    \centering
    \begin{tabular}{c|cgcccccc|c|c|c}
         Scenario & \multicolumn{7}{c}{1000 dishes} & &  SKA-like & Planck & PSD \\
         \hline
         \hline
         $N_w$ &  3 & 3 & 3 & 3& 3& 3& 3& $\bm{6}$&3&$-$&$-$ \\
         $X_\alpha$ & 3 & 3 & 3 & 3& 3& 3& $\bm{4}$& 3 &3&$-$&$-$\\
         $k^h_\text{max}\,\,\,[h\,\text{Mpc}^{-1}]$ & 0.5 &0.5 & 0.5 & 0.5 & 0.5 & $\bm{0.3}$& 0.5 & 0.5 &0.5&$-$&$-$ \\
         $k_\parallel^\text{min}\,\,\,[h\,\text{Mpc}^{-1}]$ & 0.1 &0.1 & 0.1 & 0.1 & $\bm{0.2}$ & 0.1& 0.1& 0.1&0.1&$-$&$-$\\
         $z_\text{max}$ & 12 & 12 & 12 & $\bm{10}$ & 12 & 12 & 12 & 12 & $\bm{10}$ & $-$&$-$\\
         $z_\text{min}$ &7& 7& $\bm{9}$& 7& 7& 7& 7& 7&7&$-$&$-$ \\
         $t_\text{int}$ [years] & $\bm{10}$ & 5& 5 & 5 & 5 & 5 & 5 & 5 &5&$-$&$-$ \\
         \hline
         \hline
         \rowcolor{gray!20}
         $1000\times \sigma_\tau$ & 1.9 &
         3.0 &7.0&3.8&6.9&3.8&3.9&6.3&  0.9& 7.1 & 5.8\\
         $\sigma_{M_\nu}$ [meV] &30 &33&39&33&39&33&34&38& 26&  83 & 40
    \end{tabular}
    \caption{Forecasts for the uncertainty on the optical depth and the sum of the neutrino masses for different experimental configurations. All forecasts on the optical depth include a prior from Planck and assume $f_\text{sky}=0.5$.  The neutrino mass constraints all include Planck, SO and the emission line galaxy sample of DESI (PSD) following ref.~\cite{Sailer:2021yzm}.  The baseline constraints for these are shown in the last two columns.  Columns 2-9 show the impact of varying each parameter one at a time from our fiducial to a more pessimistic scenario.  As a more futuristic example we include an SKA-like forecast, in which the collecting area is increased to $1\,\text{km}^2$, which increases the sensitivity significantly over our fiducial scenario.      }
    \label{tab:forecasts}
\end{table}

Our main results are summarized in Table \ref{tab:forecasts}, which presents forecasts for the uncertainty on the optical depth and the sum of the neutrino masses for different experimental configurations.  The 21cm-based forecasts are compared to the existing measurement precision of Planck \cite{PlanckLegacy18,PCP18} and to a forecast for how well $\tau$ and $M_\nu$ could be constrained by a combination of future CMB and galaxy redshift surveys (Planck, Simons Observatory and the emission line galaxy sample of DESI, indicated as PSD \cite{Sailer:2021yzm}).  Columns 2-9 show the impact of varying experimental configurations, one element at a time, at fixed array size ($10^3$ dishes) while the column labeled SKA-like shows the improvement afforded by a much larger array (with $1\,\text{km}^2$ collecting area), which reduces the thermal noise by roughly a factor of $5$.  We see that $z_\text{min}$, $k_\parallel^\text{min}$ and $N_w$ have the largest impact on $\sigma_\tau$, as expected from our earlier discussion.  Once $z_\text{min}>9$ we are no longer measuring the range of redshifts (or $x$) that contribute the most to $\tau$. For $k_\parallel^\text{min}>0.1\,h\,\text{Mpc}^{-1}$ we become sensitive primarily to the astrophysics controlling bubble sizes which our method explicitly marginalizes over.  When $k_\text{min}\simeq 0.2\,h\,\text{Mpc}^{-1}$ our $k_\text{max}\le k_\text{min}$ for $x>0.5$ and we are unable to probe contributions to $\tau$ from $z<8$. For larger $N_w$ the range in $\mu$ is too restricted to be able to separate the modes along the line of sight from those transverse to it, and thus we cannot break the degeneracy with bias.  By contrast we are relatively insensitive to the smallest scales we model -- changes to $X_\alpha$ and $k^h_\text{max}$ change $\sigma_\tau$ only modestly from our baseline configuration.  Column 2 and the SKA-like column shows the constraints improve non-negligibly for lower thermal noise levels, indicating that our fiducial array is far from sample-variance limited.  
The constraints from the $10\,$yr survey with our fiducial array would have the same precision as the forecasts for $\sigma_\tau$ from the proposed LiteBIRD satellite mission \cite{LiteBIRD}. In contrast to the results for a noiseless survey (\S\ref{sec:z_independent_cuts}, or SKA), here we find that a noisy survey is moderately sensitive to the maximum redshift: decreasing $z_\text{max}$ from 12 to 10 to avoid the impact of spin-temperature fluctuations can increase the uncertainty on the optical depth by $\sim25$\%. 

\section{Discussion and conclusions}
\label{sec:conclusions}

The optical depth to reionization $\tau$ is one of the key parameters in the 6-parameter $\Lambda$CDM model of cosmology \cite{PlanckLegacy18}.  Holding the amplitude of CMB fluctuation fixed, $\tau$ controls the amplitude of fluctuations in the Universe.  Currently the tightest constraints on this parameter come from observations of CMB polarization, but in the future 21cm experiments have the potential to dramatically improve upon our constraints.  We discuss an approach to constraining $\tau$ built around ideas very familiar in large-scale structure studies but less often used in reionization.  We emphasize the role of large scale modes, redshift-space distortions and a symmetries-based bias model that allows a principled but efficient marginalization over astrophysical parameters.  The split encoded in this formalism means we can focus on the cosmological information contained in large scale modes while efficiently removing sensitivity to small-scale physics (and since our expansion is based entirely on symmetry, it makes minimal assumptions about that physics). Conversely, information about the sources and morphology of reionization is encapsulated in the small-scale modes and bias coefficients, with the large-scale modes being set largely by the timing of reionization. This approach to analyzing 21cm interferometer data has implications for the range of scales and redshifts that contain most of the cosmological signal, and hence on the most impactful observational settings.

Given sufficient dynamic range in scale, and the ability to measure the anisotropy in the clustering that is imprinted by peculiar velocities, future 21cm experiments should be able to place constraints on $\tau$ that are much tighter than constraints from existing CMB experiments or even future galaxy redshift surveys and competitive with proposed, future space-based CMB anisotropy missions.  In order to make these inferences robust to uncertainties in astrophysical models, a premium should be placed on measuring large scales and on cleaning the foreground wedge as much as possible.  By contrast very high angular resolution is less helpful.  We find that much of the constraining power to $\tau$ is lost once $k_\parallel^\text{min}$ rises to $\approx 0.2\,h\,\text{Mpc}^{-1}$ or if the foreground wedge extends past $\approx 3\times$ the primary beam.  The experiment must also measure well the redshifts for which reionization is most active, i.e.\ those with $0.2<x<0.8$.

We have performed our forecasts within the context of Lagrangian perturbation theory, since this has a natural connection to the excursion set formalism and the theory of peaks and easily handles redshift-space distortions. At low redshifts the Lagrangian bias expansion has been shown to perform better for halos and galaxies than the equivalent Eulerian formalism \cite{Modi19b}, however which is preferable for reionization is currently unknown.  The Lagrangian and Eulerian approaches to perturbation theory both predict that the dynamics on the scales and redshifts relevant for determining $\tau$ are very close to linear, with small corrections that can be very well described by either formalism.  The primary limitation is not the non-linear scale but the scale of non-locality (due to ionized bubbles) in the bias expansion.  We have treated this as a fundamental limitation in our forecasts and treated the bias parameters as completely free; however, future information about the process of reionization from a combination of simulations and observations could allow us to place priors on the bias coefficients.  Such priors would allow information to be extracted from smaller scales than we have considered.  In this regard we encourage simulators (and observers) to provide information on the anisotropy of the power spectrum when publishing their models.

Our forecasts assume that reionization proceeds smoothly and in a manner that can be described as ``late and fast''.  This is consistent with the current observations, but our understanding of the process of reionization is highly incomplete.  Since the low redshift Universe is known to be highly ionized \cite{Fan06}, the main cause for concern would be multiple or partial ionizations at high $z$.  Such episodes would leave a clear signal in $P_{21}$ that would be at odds with our default scenario, so would not lead to inference of a highly precise but inaccurate $\tau$ unless $x(z)$ and the spin temperature evolution somehow conspired.  For this reason we have not considered such scenarios.

Our forecasts suggest that a scaled up version of HERA has the potential to measure the optical depth with statistical uncertainty $\sigma_\tau = 0.002-0.004$, while a SKA-like survey could achieve $\sigma_\tau = 0.0009$ (assuming a $5-10$ year integration time, we find no improvement over Planck with only $\sim1000$ hours). 
We note that this improvement in sensitivity is more modest than models which claim to accurately predict and derive $\tau$-constraints from small-scale 21cm fluctuations (e.g. refs. \cite{Liu:2015txa, Greig:2020hty, Greig:2020suk}, which use $\verb|21cmFAST|$ as the theory model).
However, these more ambitious models run an increased risk of returning biased results should their predictions differ significantly from the true small-scale 21cm signal. By contrast, our approach of modeling the 21cm intensity field using a symmetries-based bias expansion has been chosen to mitigate astrophysical uncertainties, however, several notable systematics could remain.
%
First, any error in the best-fit Weibull function to the true reionization history propagates as a systematic error to the optical depth measurement. This systematic error can be reduced by freeing up the functional form of the $x(z)$ parametrization, however, doing so (\S\ref{sec:z_independent_cuts}) comes with a significant noise penalty when a realistic foreground wedge and $z_\text{max}$ are accounted for. 
Second, we have neglected the contributions from doubly reionized helium, which is expected to contribute $\tau_\text{He}\sim 0.0008$ to the optical depth.  Even for an SKA-like survey the second ionization of helium would need to be uncertain by $\mathcal{O}(1)$ to give significant errors.  Additionally, we have neglected spin temperature fluctuations and uncertainties in the primordial helium abundance. We have shown in \S\ref{sec:forecasts} how our constraints vary with $z_\text{max}$, which can be lowered to remove any spin-temperature induced systematics, while within our formalism it would be straightforward to marginalize over any uncertainty in $Y_p$\footnote{Given a prior $\sigma_{Y_p}=0.2437\pm0.008$ from Planck \cite{PlanckLegacy18} and direct measurements of helium abundance, uncertainties in $Y_p$ would result in $\sigma_\tau\sim0.0003$. We therefore do not expect varying $Y_p$ to significantly impact our results when realistic noise is included.}.

Finally, in our discussion of perturbation theory we have made the implicit assumption that all matter (dark matter, baryons and neutrinos) in the Universe can be treated as a single fluid. In the case of neutrinos it is known \cite{Castorina2015,Bayer21} that galaxies primarily trace the cold dark matter and baryons; since the bias coefficients for the 21cm signal primarily reflect those of neutral hydrogen and ionizing sources the same approximation likely holds in this context as well. Residual differences post-recombination in the baryon ($b$) and dark matter ($c$) fields can lead to additional bias contributions proportional to relative densities and velocities \cite{Yoo11,Tseliakhovich10,Schmidt16,Chen19}. These contributions are either stationary or decay with time, and thus tend to be extremely subdominant at low redshifts, but could conceivably be a non-negligble effect at the EoR and, at $k$ beyond the acoustic scale, degenerate with the linear bias. However, we note that linear theory contributions to the observed clustering due to the relative velocity $v_{bc}$ should be significantly below 1\% of the total matter velocity contributions even at high redshifts and these terms should contribute negligibly to redshift-space distortions. Since our reionization measurement depends mostly on the observed anisotropy it should be quite immune to these additional biases.

\acknowledgments
We would like to thank Avery Meiksin for helpful discussions on Helium reionization, and Joshua Dillon for helpful discussion on thermal noise calculation. S.C.\ thanks Nickolas Kokron for useful discussions on the form of the 21cm signal.
N.S.~is supported by the Physics Division of Lawrence Berkeley National Laboratory.
S.C.\ and M.W.~are supported by the DOE.
This research has made use of NASA's Astrophysics Data System and the arXiv preprint server.
This research used resources of the National Energy Research Scientific Computing Center (NERSC), a U.S. Department of Energy Office of Science User Facility operated under Contract No.\ DE-AC02-05CH11231.

\paragraph{Note added.} While this paper was being finalized, ref.~\cite{Qin22} appeared which takes a highly related approach to modeling 21cm perturbations using a large-scale-structure-inspired bias modeling.  Our Lagrangian formalism can be exactly mapped into their Eulerian one \cite{Chen20a} so the approaches are highly synoptic.

\bibliographystyle{JHEP}
\bibliography{main}
\end{document}